\documentclass[twocolumn,prl,amsmath,amssymb,showpacs,superscriptaddress]{revtex4-1}
\usepackage{epsf}      
\usepackage{graphicx}
\usepackage{color}
\usepackage{gensymb}
\usepackage{amsmath}
\usepackage{multirow}
\usepackage{amssymb}
\usepackage{amsthm}
\usepackage{mathtools}

\begin{document}

\title{Magneto-transport and Berry phase in magnetically doped Bi$_{0.97-x}$Sb$_{0.03}$ single crystals}

\author{V. K. Maurya }
\affiliation{Department of Physics, Indian Institute of Technology Delhi, Hauz Khas, New Delhi-110016, India}
\author{Manju Mishra Patidar$^\dag$}
\affiliation{UGC-DAE Consortium for Scientific Research, University Campus, Khandwa Road, Indore 452001, India}
\author{Anita Dhaka}
\affiliation{Department of Physics, Sri Aurobindo College, University of Delhi, Malviya Nagar, New Delhi-110017, India}
\author{R. Rawat}
\affiliation{UGC-DAE Consortium for Scientific Research, University Campus, Khandwa Road, Indore 452001, India}
\author{V. Ganesan$^\ddag$}
\affiliation{UGC-DAE Consortium for Scientific Research, University Campus, Khandwa Road, Indore 452001, India}
\author{R. S. Dhaka}
 \email{rsdhaka@physics.iitd.ac.in}
\affiliation{Department of Physics, Indian Institute of Technology Delhi, Hauz Khas, New Delhi-110016, India}

\date{\today}

\begin{abstract}

We report large magnetoresistance (MR) and Shubnikov--de Haas (SdH) oscillations in single crystals of magnetically (M= Ni and Fe) doped M$_x$Bi$_{0.97-x}$Sb$_{0.03}$ ($x=$ 0, 0.02) topological insulators. The R$\bar{3}$m symmetry and phase have been confirmed by the Rietveld refinement of x-ray diffraction data. Interestingly, a magnetic field induced phase transition from semi--metallic to semi--conducting type is found with the energy gap around 80~meV at 15~Tesla in the $x=$ 0 sample. Moreover, we observe linear behavior of MR up to 15~Tesla in transverse mode and SdH oscillations in longitudinal mode where the field direction is with respect to the current and crystal plane. For the parent sample, we found the coherence length L$_\phi=$ 12.7~nm through the fitting of MR data in transverse mode with modified H--L--N equation. The extracted frequencies of SdH oscillations using the fast Fourier transform method and Landau level (LL) fan diagram are found to be consistent for the parent and Ni doped samples. The determined Fermi surface area is found to be slightly larger in Ni doped as compared to the parent sample possibly due to change in the Fermi energy. The Kohler's plot indicates a single scattering mechanism below 100~K. More importantly, the analysis with the help of LL fan diagram reveals the non-zero Berry phase $\phi_{\rm B}= -$(1$\pm$0.1)$\pi$, which demonstrates the non-trivial topological states near the Dirac point in the parent and Ni doped samples. 

\end{abstract}

\maketitle

\section{Introduction}

The discovery of quantum Hall systems demonstrated the phenomena of conducting boundaries and insulating bulk in the presence of applied magnetic field in semiconducting samples \cite{tsui, klitzig}. However, the discovery of topological insulators (TI) leads to realization of fully conducting surface states (SS) with a non-conducting bulk without any application of magnetic field due to strong spin-orbit interaction \cite{kane, hasan, moor, hsieh, qi}. In a TI, surface states are protected by the time reversal symmetry of the system and do not depend upon how the surface is cut or due to the presence of non magnetic impurities \cite{zwang}. This is because the Hamiltonian describing the SS is invariant due to small perturbations. These topological surface states were detected experimentally in Bi$_2$Se$_3$ and Bi$_2$Te$_3$ systems \cite{roushan, nishide, hsieh}. On the other hand, with the substitution of small magnetic impurities or external magnetic field results in breaking the time reversal symmetry \cite{hasan}. After the discovery of TIs, various attempts have been made in this direction, which reveal many novel and exotic phenomena as reported in refs.~\cite{dyck, dyck2005, yu, kim, maurya}. Interestingly, the appearance of bulk ferromagnetism (FM) was observed in Bi$_2$Te$_3$ with Mn substitution \cite{hor}; whereas there is no signature of FM in Fe/Mn substituted Bi$_2$Se$_3$, but a small energy gap opens at the Dirac point \cite{ychen}.

Very recently, an interesting observation by Chi {\it et al.} that thermally excited Dirac-like electrons in the L valley of the narrow gap are effective in Bi$_{1-x}$Sb$_x$ alloy to generate extremely large spin current mobility \cite{ChiSA20}. Also, Kang {\it et al.} demonstrated that BiSb thin films have a great potential for very low power spin--orbit torque switching applications \cite{Khangnm18}. Intriguingly, Li {\it et al.} have studied Nb--Bi$_{0.97}$Sb$_{0.03}$--Nb Josephson junction and showed that Dirac semimetal (DSM) opens an avenue for their applications in topological superconductivity and quantum computing \cite{LiNM18}. Moreover, a DSM to Weyl semimetal (WSM) phase transition has been reported in Bi$_{0.96}$Sb$_{0.04}$ with the application of external magnetic field \cite{kim, shin}. Also, it  can be realized through negative longitudinal magnetoresistance (LMR), which is a physical demonstration of chiral anomaly effect and leads to violation of Ohm's law in the bulk transport measurements, which originated from the non-trivial topological structure \cite{shin}. In fact first principles calculations show possible WSM states even at higher concentration of Sb (50\% and 83\%) in the specific inversion symmetry broken conditions \cite{su}. Interestingly, a series of Bi-Sb samples have been investigated by bulk magneto-transport measurements where 16\% Sb substituted sample show a strong negative LMR and non-linear I-V characteristics \cite{amit}. Further, note that around 3\% Sb substitution at Bi site causes the closing of the gap; whereas at higher concentration, the gap opens up due to the realization of 3D Dirac points and the system behaves as direct gap insulator and its low temperature properties are defined as spin-orbit coupled Dirac particle \cite{fu, TeoPRB08}. The coexistence of electron and hole pockets has been reported in Bi-Sb and other Weyl semimetals \cite{mangez, fauque, heremans, yang}. Sasaki {\it et al.} used quantum oscillation measurements and showed the possibility of chiral edge mode in Bi--Sb from the SS in the presence of magnetic fields \cite{Sasaki}. Note that there are reports on the magnetic ion substituted topological insulators like Bi$_2$Se$_3$, Bi$_2$Te$_3$ \cite{dyck, dyck2005, yu, kim, maurya}; however, magnetically doped Bi-Sb crystals have largely been unexplored \cite{AfzalJSNM20}. Therefore, it is vital to investigate the MR behavior of Bi-Sb and the effect of magnetic-ion substitution on the transport properties. 

In order to study the dispersion relation in a TI near Dirac point, Shubnikov de Haas (SdH) oscillations play an important role where its phase factor connects to the Berry phase \cite{taskin}. Here, a non trivial nature of the Berry phase (a $n_x\pi$ Berry phase) indicates that the charge carriers showing oscillations are Dirac fermions \cite{taskin}. On the other hand, a trivial Berry phase (zero Berry phase) shows the ordinary charge carriers movement. To get the exact value of the Berry phase, we need to plot the Landau level (LL) fan diagram where the LL index $n$, obtained from the correctly assigning the maxima/minima in $\rho_{xx}$ data, is plotted against 1/B \cite{taskin}. The frequency of SdH oscillations can be extracted by the slope of the straight line fit ($n$ vs 1/B curve) as well as an independent analysis using fast Fourier transform (FFT). The intercept on the $x-$axis (when 1/B = 0), $n_x$ gives the information about the trivial or non trivial Berry phase. Interestingly, in Bi$_{1-x}$Sb$_x$ a topological phase transition has been reported at $x=$ 0.03--0.04 where the presence of magnetic field can change the state from Dirac semimetal to Weyl fermions \cite{hasan, fu, TeoPRB08, kim, shin}. Therefore, in this paper we present a detailed investigation of magnetoresistance and SdH oscillations in the parent and 2\% doped (Ni and Fe at the Bi site) Bi$_{0.97}$Sb$_{0.03}$ single crystals. We observe magnetic field induced suppression of the semi--metallic state in the parent sample with the energy gap of about 80~meV at 15~Tesla. Our focus in these low magnetically doped Bi-Sb is to break time reversal symmetry without altering the bulk state as the FM state can only alter the surface state. Our thorough analysis of SdH oscillations by plotting the LL fan diagram reveal a non-zero Berry phase, which suggests for non-trivial topological states near Dirac point in these samples.

\section{Experimental details}

Single crystals of Bi$_{0.97}$Sb$_{0.03}$, Ni$_{0.02}$Bi$_{0.95}$Sb$_{0.03}$ and Fe$_{0.02}$Bi$_{0.95}$Sb$_{0.03}$ were prepared by vertical modified Bridgman furnace. The stoichiometric ratio of Bi, Sb, Ni/Fe (all $\ge$99.9\% purity) are sealed in an evacuated quartz tubes and heated up to 650$^{\rm o}$C for 5 days, then cooled to 350$^{\rm o}$C during the period of 3 days. The samples were sintered at 350$^{\rm o}$C for 4 days and finally quenched in cold water. We got silvery crystals by cleaving with the help of a scalpel blade. These crystals/powder samples were characterized by x-ray diffraction measurements using a diffractometer from RIGAKU (Miniflex 600 model) and the data were analyzed by GSAS software using the Rietveld refinement method. The magnetoresistance and transport measurements were performed in four probe arrangement at UGC-DAE CSR, Indore using a home-built setup attached with the Oxford superconducting magnet. Also, a Quantum design PPMS was used at UGC-DAE CSR, Indore for the measurements up to 15~Tesla field. 
	     
  \section{Results and discussion}
  
    \begin{figure}
 \centering
  \includegraphics[width=3.4in]{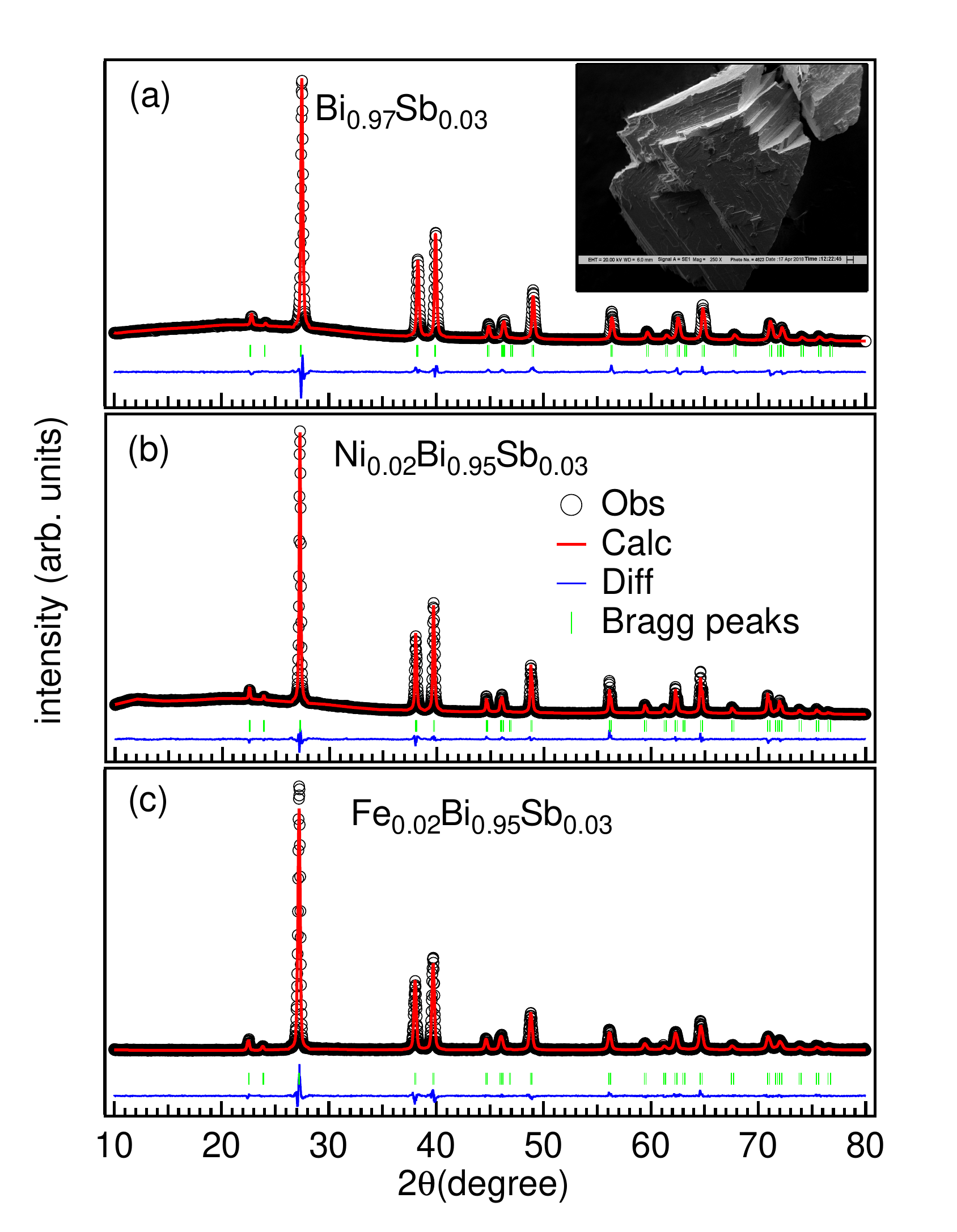} 
\caption{(a--c) The powder x-ray diffraction patterns of all the three samples along with the Rietveld refinement. The inset in (a) shows SEM image of a crystal chunk.}
\label{XRD}
\end{figure} 

\begin{figure}
 \centering
  \includegraphics[width=3.0in]{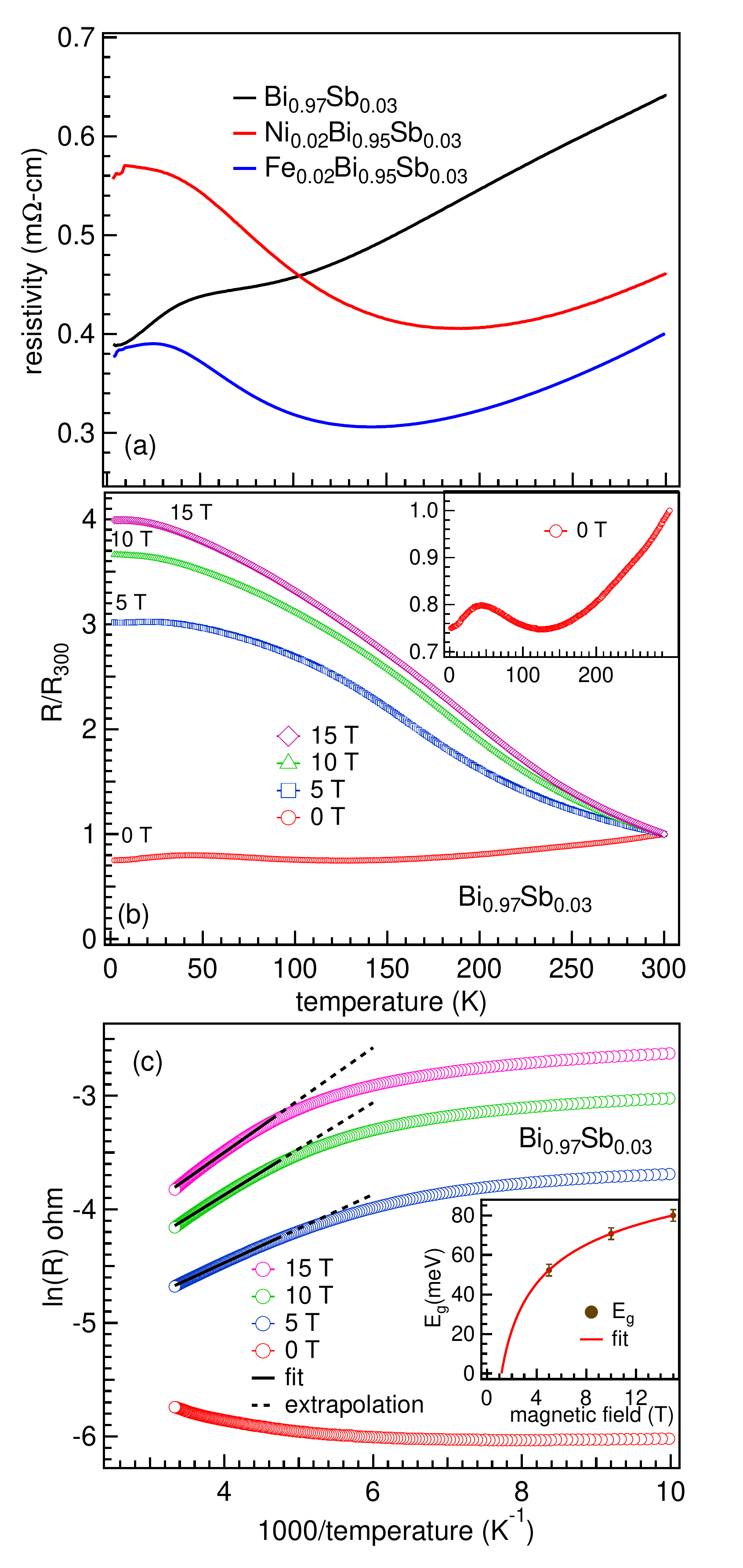}
\caption{(a) The dc-electrical resistivity curves of all the three samples. (b) The resistance of Bi$_{0.97}$Sb$_{0.03}$ plotted as a function of temperature measured in B $\perp$ crystal plane configuration. The inset in (b) shows the behavior at zero field. (c) The fitting with Arrhenius equation (black line) of the data (open circles) in the high temperature range (215--300~K). The inset in (c) shows the variation of energy gap with the applied magnetic field along with the power law fit.}
\label{resistivity}
\end{figure} 
  
  Figs.~1(a--c) show the powder x-ray diffraction (XRD) patterns and the Rietveld refinement using GSAS software confirm the phase purity and R$\bar{3}$m symmetry of the parent Bi$_{0.97}$Sb$_{0.03}$ as well as both the doped samples. The determined cell parameters are $a=b=$ 4.5~\AA, $c=$ 11.8~\AA~ and the cell volume is 211.45~\AA$^3$ for these samples, which are in good agreement with reported in refs.~\cite{Sultana, SinghPRB16}. The XRD data of freshly cleaved crystals show only (00{\it l}) reflections (not shown) except a small (202) peak at 2$\theta$$\approx$ 49$\degree$ \cite{Sultana}. Further, the surface morphology has been characterized using scanning tunneling microscopy (SEM), which confirms the layered growth of the sample, see the inset in Fig.~\ref{XRD}(a). We have also checked the stoichiometric composition of these samples using energy dispersive x-ray spectroscopy (not shown). 
  
 \begin{figure*}
\centering
\includegraphics[width=7.15in]{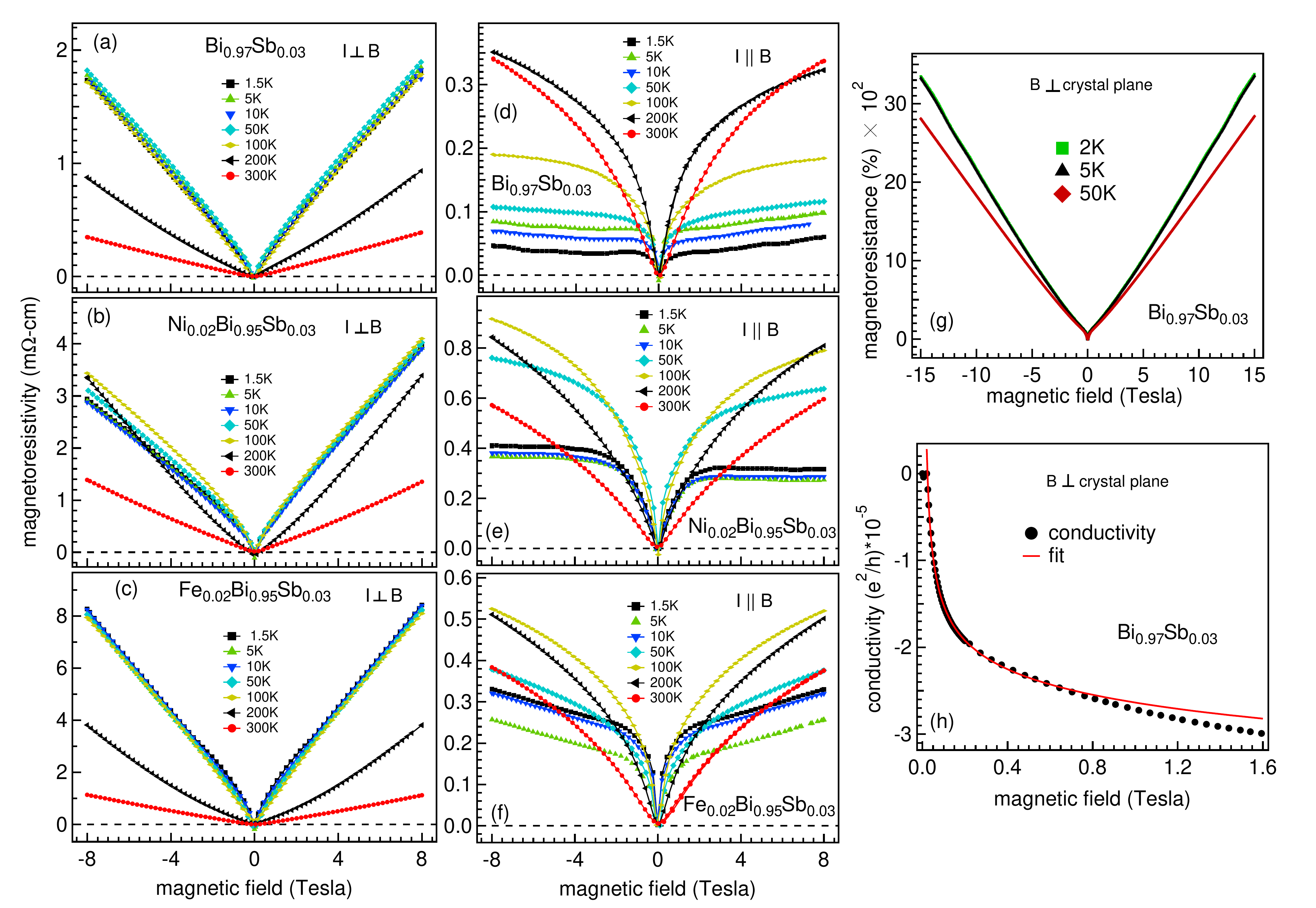}
\caption{The magnetoresistivity [$\rho$($\rm {B}$)$-\rho$(0)] curves measured when applied magnetic field is in the crystal plan, but perpendicular to the current in (a--c) and parallel to the current flow in (d--f). The MR(\%) measured at few temperatures in the transverse magnetic field (up to 15~T) direction with respect to the crystal plane as well as current flow (g). The magneto conductivity data fitted with modified H--L--N equation in the low field region (h).} 
\label{per}
\end{figure*} 
  
The resistivity behavior of these samples is compared in the Fig.~\ref{resistivity}(a) where in case of the parent sample the resistivity decreases with cooling, which is a signature of a typical topological insulator type semimetal behavior. We observe the decrease in the resistivity values at the room temperature with the substitution of Fe and Ni at the Bi site. In case of both the doped samples the interesting observation is that the resistivity decreases initially with cooling, but below about 150~K there is a strong upturn in the resistivity unlike the parent sample. This behavior of resistivity can be expected when the contribution is from the competing nature of two different sources known as bulk and surface transport channels. Therefore, at higher temperatures a strong semi-metallic behavior dominate due to bulk conduction; whereas at low temperatures the current starts flowing through surface \cite{nishide, amit}, which reflects in higher bulk resistivity below 100--150~K. Moreover, we have measured the temperature dependent resistance of Bi$_{0.97}$Sb$_{0.03}$ sample measured in B $\perp$ to $ab$ plane (crystal plane) configuration at various fields, as shown the normalized curves in Fig.~\ref{resistivity}(b). In the inset we show the behavior without any field (0 T) where the resistance decreases monotonically with decrease in temperature like a semimetal. On the other hand, when the magnetic field value increases to 5~T, the resistance changes to semi-conducting like behavior, which remains up to the highest measured field value of 15~T. Further, the resistance values increase significantly and saturate below $\approx$20~K temperatures measured at high magnetic fields. Interestingly, the observed phase transition from semi-metallic to semi-conducting type indicates an increase in the energy gap due to change in the band structure with the magnetic field \cite{YueAPL15}. Notably similar type of phase transition has also been reported in graphite, Bi \cite{DuPRL05, YangScience99} as well as in transition metal dichalcogenide WTe$_2$ \cite{AliNature14} and topological Weyl semimetals NbP \cite{ShekharNPhys15}. Moreover, in Fig.~\ref{resistivity}(c) we show the plot of ln(R) as a function of inverse temperature for the data measured at 0, 5, 10 and 15~Tesla fields. In order to find the gap size we fit the data between 215--300~K using a simple thermal activation type Arrhenius equation: R(T) $\propto$ exp(E$_g$/2k$_B$T) where R is the resistance and E$_g$ is the bandgap energy. In the inset of Fig.~\ref{resistivity}(c) we plot the determined E$_g$ values at different fields, which clearly shows increasing behavior with increasing the magnetic field \cite{YueAPL15}. Though there are three data points, the power law fit reveals that the energy gap opens at $\ge$1~Tesla field in Bi$_{0.97}$Sb$_{0.03}$ and the value of E$_g$ is found to be around 80~meV at 15~Tesla. The energy gap was reported in Bi--Sb system at low temperature \cite{hsieh}, with Sb doping \cite{LenoirJPCS96}, and with film thickness \cite{HiraharaPRB10}. 

\begin{figure*}	                          	                    
\centering   	                          	                    	         
\includegraphics[width=7.2in]{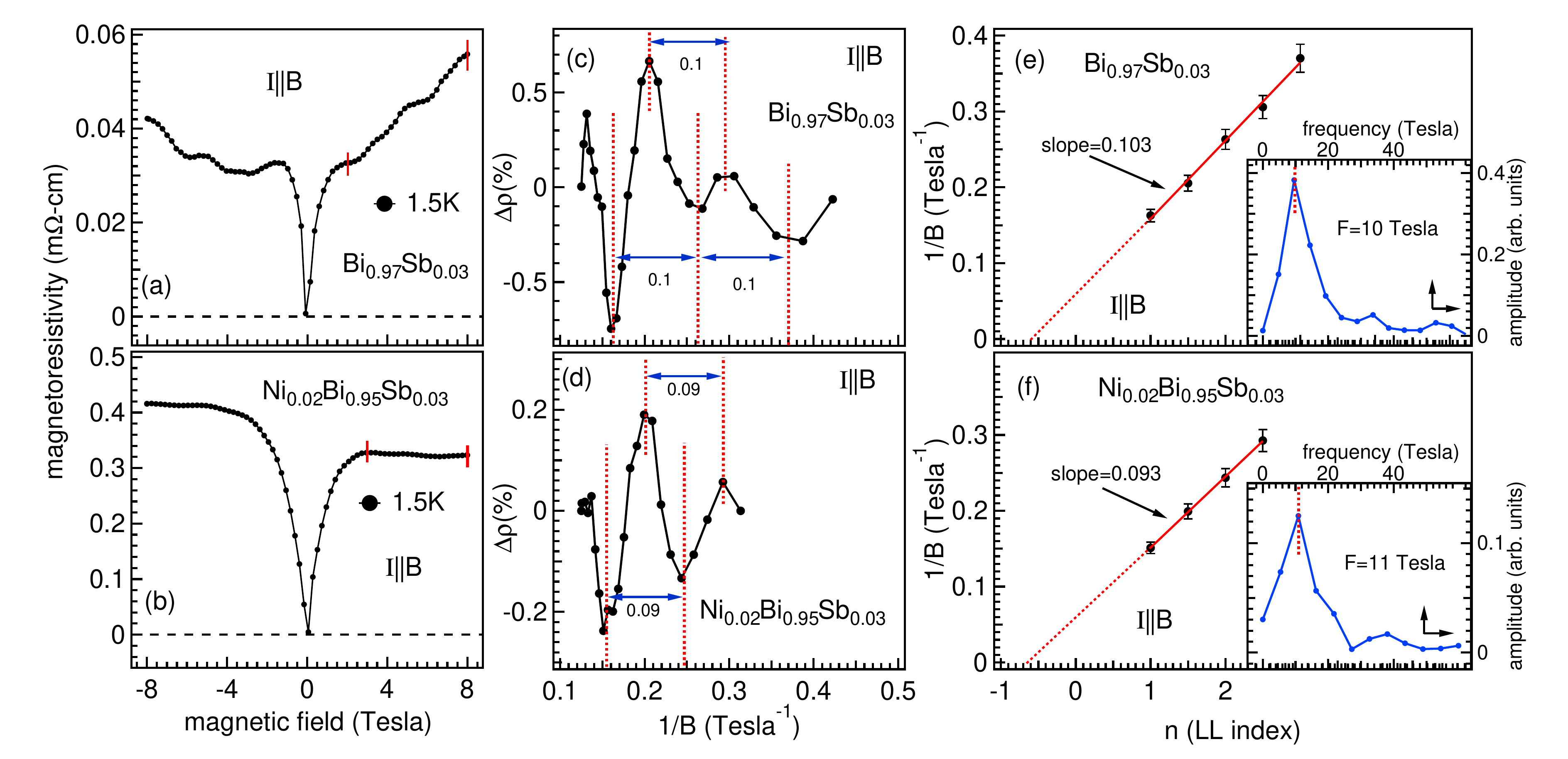}
\caption{(a, b) The MR curves (SdH oscillations) measured in I$\parallel$B configuration at 1.5~K. (c, d) The SdH oscillations resolved after subtracting the baseline constructed by a polynomial background from the MR data, periodicity of minima is shown by red dashed lines. (e, f) The LL fan diagrams (1/B vs $n$) are shown where the intersect at $x-$axis gives the value of $n=$ --0.5$\pm$0.1 and inverse of slope gives the value of frequency for both Bi$_{0.97}$Sb$_{0.03}$ and Ni$_{0.02}$Bi$_{0.95}$Sb$_{0.03}$ samples, respectively. The FFT plots for the SdH oscillations are shown in the insets of (e, f) as a function of F(T) frequency for both the samples.}
\label{SDH}       	         	                          	                    
\end{figure*} 

In Figs.~\ref{per}(a--f), we plot the magnetoresistivity (MR) data at various temperatures between 300~K and 1.5~K where the current (I) and applied magnetic field (B) are in the $ab$ plane of the crystals Bi$_{0.97}$Sb$_{0.03}$, Ni$_{0.02}$Bi$_{0.95}$Sb$_{0.03}$ and Fe$_{0.02}$Bi$_{0.95}$Sb$_{0.03}$, respectively. The data in Figs.~\ref{per}(a--c) are measured in I$\perp$B configuration; whereas I$\parallel$B configuration was used for the data presented in Figs.~\ref{per}(d--f). In case of I$\perp$B, we observe almost linear and non-saturating behavior of MR up to 8~Tesla magnetic field, which is conventional for TIs as also reported in literature \cite{YueAPL15}. There is no change in the magnitude of MR between 1.5~K and 100~K; however, above 100~K, the MR decreases significantly for all the samples. Moreover, for the Ni and Fe substituted samples, an increase in MR magnitude is appreciable at all temperatures and applied magnetic fields, as shown in Figs.~\ref{per}(a--c). Unlike I$\perp$B configuration, the MR curves in Figs.~\ref{per}(d--f) are non-linear in nature and overall the magnitude is much smaller at all measured temperatures for these samples when compared in Figs.~\ref{per}(a--c). First we note that all the MR curves show a cusp like minimum below about 1~Tesla and up to 100~K. This dramatic increase in MR in low magnetic field range and then becoming linear clearly indicate the presence of weak antilocalization in these samples, which can be attributed due to 3D Dirac fermions \cite{HLN}. On the other hand, above 200~K, the MR dip at low B is broadened significantly and almost disappeared $\ge$ 300~K due to the decrease in the phase coherence length at high temperatures. Interestingly, in the I$\parallel$B configuration there is no significant increase in the MR values at low temperature. Also, we did not observe negative MR either in parallel or perpendicular field direction to the applied current. It is worth to note that Boer {\it et al.} observed the negative MR in Bi-Sb due to the chiral magnetic effect \cite{BoerPRB19}. 

In order to further understand the behavior at much high fields, we have measured the transverse field MR data for parent sample Bi$_{0.97}$Sb$_{0.03}$, as shown in Fig.~\ref{per}(g) at 2~K, 5~K and 50~K temperatures. In this case, the applied magnetic field is in the transverse direction with respect to the current flow as well as the crystal plane. The percentage of MR is calculated with the equation $\Delta\rho /\rho(0)=100\times[\rho(\rm {B})-\rho(0)]/\rho(0)$. The magnitude of MR\% is of the order of 3000 at around 15~Tesla field, which found to be higher than the other configuration (field parallel to crystal plane). This large magnitude of MR\% also indicates high quality of the crystals \cite{Sultana, YueAPL15, KozhemyakinJAP17}. Further, we have performed magneto conductivity analysis using Hikami-Larkin-Nagaoka (H-L-N) model \cite{HLN, AdroguerPRB15, SultanaJSNM18}, which gives the information of contribution from the sheet surface conduction as compared to the bulk conduction in the total conductivity and a qualitative analysis of weak antilocalization (WAL) effect. Therefore, the conductivity is plotted at 1.5~K in Fig.~\ref{per}(h) and fitted in the low field region according to the H--L--N equation as written below \cite{HLN, AdroguerPRB15,SultanaJSNM18}:
       \begin{equation}
        \Delta\sigma = \frac{\alpha e^2}{\pi h} \left[\psi \left(\frac{h}{8 \pi e B L_\phi^2} + \frac{1}{2} \right) - \ln \left( \frac{h}{8 \pi e B L_\phi^2} \right)\right] 
\end{equation}
where $\Delta \sigma$ denotes the change in magneto conductivity, $\alpha$ term takes into account for the effective channels contributing in conduction, $\psi$ is digamma function, and L$_\phi$ represents the coherence length. We have tried to fit the conductivity using eq.~1; however, the data could not be fitted properly (not shown) and the obtained value of $\alpha$ is significantly larger. Therefore, we use an additional term $\beta$/B to take into account for the classical (linear) MR effect \cite{WangNanoR15}. In Fig.~\ref{per}(h) we show the fitting of the conductivity curve using above eq.~1 plus an additional term ($\beta$/B), in transverse magnetic field configuration. We found the values of fitting parameters like L$_\phi=$ 12.7~nm and $\alpha=$ -1.1$\times$10$^{-5}$, where an almost zero value for $\alpha$ indicates that the conduction is purely 2D in the absence of magnetic impurity because bulk conductivity is filtered out by the $\beta$ term. The low value of $\beta=$ 0.03 implies a very poor contribution of the bulk to the conductivity. 
	           
Note that there is a strong dependence of MR signal on the temperature and field, as shown in Fig.~\ref{per}(d--f). Therefore, we further analyzed these data measured in I$\parallel$B configuration for the Shubnikov-de Haas (SdH) oscillations. In this context, we have plotted in Fig.~\ref{SDH}(a) the MR data of Bi$_{0.97}$Sb$_{0.03}$ measured in I$\parallel$B configuration at 1.5~K, which clearly show the SdH oscillations at fields above 2~Tesla \cite{TaskinPRB10}. These SdH oscillations are suppressed in Ni substituted sample, but weak oscillations are clearly visible in the inset plot, as shown in Fig.~\ref{SDH}(b). On the other hand, in case of Fe substituted sample the SdH oscillations completely suppressed (not shown). This suppression is probably due to the breaking of time reversal symmetry with doping of magnetic elements or increased disorder with doping \cite{VuPRB19}. As the SdH oscillations are clearly visible for the Bi$_{0.97}$Sb$_{0.03}$ sample, we have fitted a baseline and subtracted the polynomial type background from the data (between 2 and 8~Tesla field), marked by vertical red lines in Fig.~\ref{SDH}(a). 
 \begin{figure*} 
\centering
\includegraphics[width=6.9in]{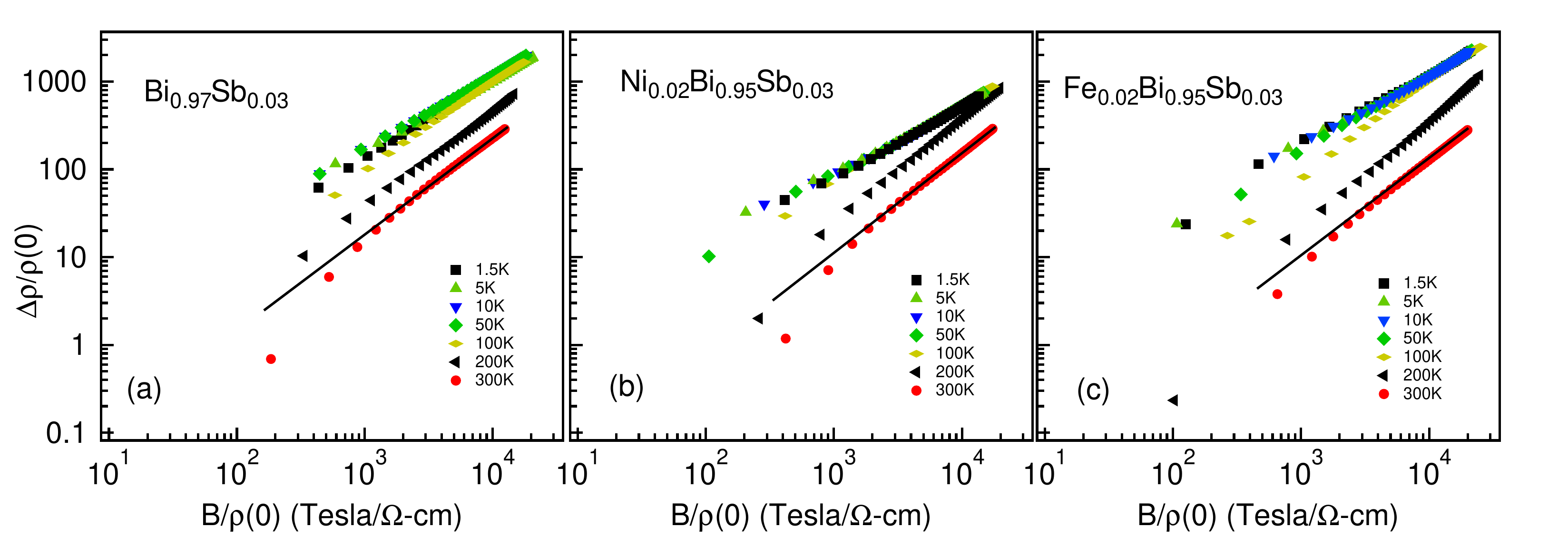}
\caption{Log plot of reduced resistivity $[\Delta \rho/ \rho(0)]$ (in I$\perp$B configuration) against the reduced magnetic field with Kohler scaling. } 
\label{kohler}
\end{figure*}
The resulting plot is shown in Figs.~\ref{SDH}(c), which clearly shows 2-3 oscillations having 1/B periodicity ($\approx$0.1~Tesla$^{-1}$) \cite{BanerjeePRB08}, which confirm the well defined Fermi surface (FS). Therefore, for this sample, in the inset of Fig.~\ref{SDH}(e) we present the fast Fourier transform (FFT) of the oscillations performed with the sampling interval 0.008, which shows a single prominent peak at F$_{\alpha}=$ 10$\pm$0.3~Tesla frequency \cite{LiNM18}. The FFT analysis of the SdH oscillations can help to resolve the number of Fermi pockets involved in the FS. We can calculate the cross sectional area (A$_f$) of the FS using the Onsager relation F = ($\phi_0$/2$\pi^2$)A$_f$, where F is frequency, $\phi_0= h/2e$ is defined as the magnetic flux quantum (2.068$\times$10$^{-15}$~Wb), $h$ is the Planck constant and $e$ is electron charge. The obtained A$_f$ value is about 9.53$\times$10$^{12}$ cm$^{-2}$ corresponding to the F$_{\alpha}$ peak, which is indeed a very small fraction of the total FS, which represents the area enclosed by electron's cyclotron motion. The Fermi wave vector [k$_f=$ (A$_f/\pi$)$^{1/2}$] corresponds to the A$_f$ value is found to be k$_f=$ 1.74$\times$10$^{6}$ cm$^{-1}$,which is consistent with the Fermi wave number values extracted from the FS measured by angle-resolved photoemission spectroscopy (ARPES) \cite{HiraharaPRB10}. Also, using the magneto-infrared spectroscopy, Schafgans {\it et al.} found a multitude of Landau level transitions in the Bi$_{0.91}$Sb$_{0.09}$ sample and the k$_f$ of shortest FS ellipsoid was found to be 2.3$\times$10$^{5}$ cm$^{-1}$ \cite{SchafgansPRB12}, which is close to the value (4.15$\times$10$^{5}$ cm$^{-1}$) determined by Taskin {\it et al.} on the same composition in refs.~\cite{taskin, TaskinPRB10}.  Also, we have calculated the surface charge carrier concentration using $n_s=$ k$_f^2$/4$\pi$, which is found to be 2.4$\times$10$^{11}$ cm$^{-2}$ for the Bi$_{0.97}$Sb$_{0.03}$ sample. Taskin {\it et al.} reported the value of $n_s=$1.4$\times$10$^{10}$ cm$^{-2}$ for 2D FS of Bi$_{0.91}$Sb$_{0.09}$ sample \cite{taskin}, which almost one order of magnitude different than the observed for the Bi$_{0.97}$Sb$_{0.03}$ sample in this study. Also, the electron concentration of $n=$ 1.8$\times$10$^{17}$ cm$^{-3}$ (Bi$_{0.91}$Sb$_{0.09}$) was reported in ref.~\cite{taskin} by the Hall measurements and a carrier density of $n=$ 4$\times$10$^{16}$ cm$^{-3}$ (Bi$_{0.96}$Sb$_{0.04}$) in ref.~\cite{BanerjeePRB08} by magneto-transport study. Moreover, for the Ni$_{0.02}$Bi$_{0.95}$Sb$_{0.03}$ sample, the SdH oscillations are weakly visible ($\approx$0.09~Tesla$^{-1}$ periodicity); therefore, we have analyzed the data (between 2 and 8~Tesla field), marked by vertical red lines in Fig.~\ref{SDH}(b), in similar fashion and plotted in Figs.~\ref{SDH}(d, f). Here, the calculated A$_f$ value is 1.05$\times$10$^{13}$ cm$^{-2}$ corresponding to the F$_{\alpha}=$ 11$\pm$0.3~Tesla frequency [see the FFT in the inset of Fig.~\ref{SDH}(f)]. Further the Fermi wave vector and surface carrier concentration values are found to be about k$_f=$ 1.83$\times$10$^{6}$ cm$^{-1}$ and n$_s=$ 2.7$\times$10$^{11}$ cm$^{-2}$, respectively. The carrier concentration values are relatively smaller, which indicate that the Fermi level is close to the Dirac point of the topological SS \cite{ShresthaPRB14}. Our analysis indicates a slight increase in the FS cross section area with Ni substitution as compare to the parent sample. This is possibly due to the fact that the FS passes through the Dirac point for Bi$_{0.97}$Sb$_{0.03}$; whereas the Fermi energy would increase with Ni substitution, which may result in larger FS area. Note that Guo {\it et al.} and Nakamura {\it et al.} used ARPES to measure the SS across the topological phase transition, which found an expansion of the bandwidth due to reduction of the spin-orbit coupling in Bi$_{1-x}$Sb$_x$ \cite{GuoPRB11, NakamuraPRB11}.

It is important to note here that the SdH oscillations can be observed mostly in a 2D electron gas condition, which in the present case fulfil due to conducting channels at the surface. In the presence of magnetic field, electrons in  the  bulk  region perform a circular motion. Nevertheless, the electrons near the  edge of the sample cannot perform full circular motion, because  of the back scattering from the boundary of the sample. Therefore, these electrons at the surface make the conduction possible due to their half circular motion on the Fermi surface. The resulting energy bands are called Landau levels (LL) and the gap between the LL increases with the magnetic field. To understand further the origin of SdH oscillations and the nature of band dispersion near Dirac point, we have extracted the LL fan diagram and the plots of 1/B versus Landau index ($n$) are shown for both the samples in the insets of Figs.~\ref{SDH}(e, f), respectively, where $n$ assigned to the maxima/minima in the SdH oscillations \cite{ZhaoNPJQM, ZhaoPRB19}. Here, the correct assignment of $n$ to either the maxima or minima in SdH oscillations is according to the conductance matrix and crucial to avoid an extra factor in $\gamma=$ 1/2 \cite{XiongPRB12}. As in the present case $\rho_{xx} \ll \rho_{xy}$ and normally the bulk conduction dominates in TIs, we have assigned the integer/half-integer index to the local maxima/minima of the SdH oscillations \cite{XiongPRB12}, as presented in Figs.~\ref{SDH}(c, d). Ideally, a straight line fit to the 1/B versus $n$ plot should cut at some $n_x$ value where 1/B $\to$ 0. Here, the $n_x$ value is found to be --0.5$\pm$0.1 for both the samples, which suggests the $\pi$ shift related to the Berry phase of surface electrons \cite{XiongPRB12}. It is important to note that the inverse of slope of the linear fitting leads to the frequency F = 9.7~Tesla and 10.7~Tesla for both the samples, respectively, which are very close to the frequency values determined by FFT in the insets of Figs.~\ref{SDH}(e, f) and validates the analysis. Moreover, in order to get insight of topological nature of these samples, the value of $\gamma$ can be determined using $\gamma = 1/2 -n_x$, and the Berry phase ($\phi_{\rm B}$) can be calculated by the relation $\phi_{\rm B}=\pi(1-2\gamma)$. The value of the Berry phase decides the nature of topological bands either trivial ($\phi_{\rm B}=$ 0) or non-trivial ($\phi_{\rm B}=\pi$) \cite{MikitikPRL99}. In the present case, a non zero value of $\phi_{\rm B}= -($1$\pm$0.1)$\pi$ suggests for non-trivial phase \cite{ZhangPRB09}, which corresponds to the Dirac fermions and demonstrate a linear band dispersion in both the samples \cite{BoerPRB19, BaruaJPCM15}. Notably the non-trivial nature of bands has been realized in Bi-Sb alloys using first-principles calculations \cite{SinghPRB16,ZhangPRB09}. This is consistent as the Berry phase associated with the helical SS, which should result in weak antilocalization in the electron transport \cite{fu} arising from either spin-momentum locking in the topological SS and/or strong spin-orbit interaction in the bulk. A surface electron moving in the circles above the Dirac point could acquire a $\phi$ Berry phase. The value of $\phi_B$ for a nominal magnetically doped TI suggests that magnetic states and non trivial topological states can coexist if we optimise the doping such that it does not affect the bulk states. As shown in Fig.~\ref{per} and discussed before, the rapid increasing behavior of MR when breaking time reversal symmetry with small applied field can be ascribed to WAL, which occurs during the diffusive transport of electrons in the quantum interference effects, and the phase coherence lengths are found to be larger than the mean free path. These observation are consistent with the Adler-Bell-Jackiw anomaly and the Weyl-metal picture \cite{KimPRL13}. The WAL has been reported in many topological samples including Bi$_{1-x}$Sb$_x$ single crystals maintaining Dirac fermions due to the $\pi$ Berry phase appeared from circular motion around the FS \cite{VuPRB19}. This $\pi$ Berry phase then induces destructive quantum interference, which can suppress backscattering of electrons and lead to an enhancement in the conductivity with lowering the sample temperature. Also, the Fermi surface measured using ARPES confirms the non-trivial Z$_2$ topological number in Bi$_{1-x}$Sb$_x$ films \cite{HiraharaPRB10}.

Finally, in order to study the role of scattering due to magnetic doping we present the analysis of the MR data measured in I$\perp$B configuration by fitting with the Kohler's equation $\Delta\rho/\rho(0) = A[{\rm {B}/\rho(0)}]^b$. The MR curves taken at various temperatures can be correlated by doing the Kohler's scaling of the magnetic field with the resistance at zero field. We note that the product of the cyclotron frequency (which depends on the magnetic field and is independent of temperature) and the scattering time (which depends on the temperature) determine the temperature dependent MR data. According to the Kohler's rule, all the curves of MR plotted against inverse magnetic field scaled by $\rho(0)$ at different temperature should merge into a single curve. In the logarithmic scale as shown in the Fig.~\ref{kohler}, different slope of curves (value of $b$) represents a change in the scattering process at that particular temperature, which can be either in the form of sign change, mobility or density of charge carriers. For all the samples, the curves below 100~K merges; whereas the slope of the Kohler's plots changes at higher temperatures, which point out the possibility of single scattering mechanism. We found that the value of $b$ varies roughly between 0.8 and 1.2 for in these samples. 

\section{\noindent ~Conclusions}

We have studied the magnetoresistance (MR) and SdH oscillations in magnetically doped Bi-Sb topological insulators using magneto-transport measurements in large temperature range. The resistance shows a typical semi--metallic behavior, which changes to semi--conducting type at high magnetic fields where the energy gap opens at $\ge$1~Tesla and is found to be around 80~meV at 15~Tesla. The MR data measured at 1.5~K show the quantum oscillations when the current direction is parallel to the applied magnetic field. The Kohler's plot indicates a single scattering mechanism below 100~K in all the samples. The MR data in transverse configuration and fitting using modified H--L--N equation in the low field region gives the coherence length L$_\phi=$ 12.7~nm for the parent sample. Our detailed analysis confirms magnitude and number of oscillations, and the fast Fourier transform (FFT) and LL fan diagram determine the frequency (F) of the oscillations, which found to be consistent for both parent and Ni substituted samples. The determined cross sectional Fermi surface area is found to be A$_f=$ 9.53$\times$10$^{12}$ cm$^{-2}$, and the Fermi wave vector and concentration values are k$_f=$ 1.74$\times$10$^{6}$ cm$^{-1}$ and $n_s=$  2.4$\times$10$^{11}$ cm$^{-2}$, respectively, determined from the frequencies of SdH oscillations for the parent sample. Similarly, for the Ni doped sample, we obtain the values of $A_f=$ 1.05$\times$10$^{13}$ cm$^{-2}$, and the k$_f=$ 1.83$\times$10$^{6}$ cm$^{-1}$ and the $n_s=$ 2.7$\times$10$^{11}$ cm$^{-2}$ for the Ni$_{0.02}$Bi$_{0.95}$Sb$_{0.03}$ sample. The FS area is found to be slightly larger in Ni doped as compared to the parent sample, which possibly due to the change in the Fermi energy. Interestingly, the calculated Berry phase with the help of LL fan diagram is non-zero, which demonstrate non trivial topological state and transport phenomena through the Dirac point in these samples.     

\section{\noindent ~Acknowledgments}

VKM thanks SERB-DST for the fellowship through NPDF scheme (no. PDF/2017/001695).  Authors thank UGC-DAE CSR, Indore centre for detailed magneto-transport measurements. We acknowledge the physics department and central research facilities (CRF), IIT Delhi for XRD, SEM and EDX data. VKM also thanks Mr. Sachin Kumar for help during the measurements. RSD gratefully acknowledges the financial support from BRNS through DAE Young Scientist Research Award project sanction No. 34/20/12/2015/BRNS.

\end{document}